# The Common Origin of High-energy Protons in Solar Energetic Particle Events and Sustained Gamma-ray Emission from the Sun

N. Gopalswamy

Code 671, Solar Physics Laboratory, NASA Goddard Space Flight Center, Greenbelt, MD 20771

nat.gopalswamy@nasa.gov

S. Yashiro[1], P. Mäkelä[1], H. Xie[1], and S. Akiyama[1]

Department of Physics, The Catholic University of America, Washington, DC 20064

**Short Title: Common Origin of High-energy Protons in SEP and SGRE Events**

[1]also at NASA Goddard Space Flight Center, Greenbelt, MD 20771

## ABSTRACT

We report that the number of >500 MeV protons (Ng) inferred from sustained gamma-ray emission (SGRE) from the Sun is significantly correlated with that of protons propagating into space (NSEP) as solar energetic particles (SEPs). Under the shock paradigm for SGRE, shocks driven by coronal mass ejections (CMEs) accelerate high-energy protons sending them toward the Sun to produce SGRE by interacting with the atmospheric particles. Particles also escape into the space away from the Sun to be detected as SEP events. Therefore, the significant $N_{SEP} - N_g$ correlation (correlation coefficient 0.77) is consistent with the common shock origin for the two proton populations. Furthermore, the underlying CMEs have properties akin to those involved in ground level enhancement (GLE) events indicating the presence of high-energy (up to ~GeV) particles required for SGRE. We show that the observed gamma-ray flux is an underestimate in limb events (central meridian distance >60º) because SGRE sources are partially occulted when the emission is spatially extended. With the assumption that the SEP spectrum at the shock nose is hard and that the 100 MeV particles are accelerated throughout the shock surface (half width in the range 60º – 120º) we find that the latitudinal widths of SEP distributions are energy dependent with the smallest width at the highest energies. Not using the energy-dependent width results in an underestimate of $N_{SEP}$ in SGRE events occurring at relatively higher latitudes. Taking these two effects into account removes the apparent lack of $N_{SEP} - N_g$ correlation reported in previous studies.

**Subject headings**: Sun: coronal mass ejections - Sun: flares - Sun: particle emission - Sun: gamma-ray emission - shock waves

# 1. Introduction

Gamma-ray emission from solar eruptions can last for minutes to many hours beyond the impulsive phase of the associated flare (Forrest et al. 1985). The gamma-ray spectrum of the 1982 June 3 event observed by the Solar Maximum Mission's Gamma-ray Spectrometer indicated that neutral pions are created during both impulsive and post-impulsive phases of the eruption due to the precipitation of >300 MeV protons to the chromosphere. Forrest et al. (1985) suggested that the post-impulsive-phase emission may be a signature of the acceleration process which produces solar energetic particles (SEPs) in space. Early theoretical investigations indicated that the impulsive-phase and the extended-phase emissions can be explained by different populations of energetic protons with different spectral characteristics (Murphy et al., 1987): stochastic spectrum from the flare reconnection region and shock spectrum from the front of CME-driven shocks. Shocks survive for long distances into the heliosphere, sometimes reaching the edge of the solar system (Richardson et al. 2005), so there is continued acceleration of energetic particles that can naturally explain the longevity of the post-impulsive-phase gamma-ray emission. Under the flare-particle paradigm, the longevity of the post-impulsive-phase gamma-ray emission is explained by the trapping of flare particles in flare loops with suitable presence of magnetohydrodynamic turbulence (Ryan and Lee, 1991). The post-impulsive-phase gamma-ray emission was originally referred to as long-duration gamma-ray flare (LDGRF, e.g., Chupp and Ryan 2009) emphasizing the flare-origin of the underlying energetic protons. With the advent of the Large Area Telescope (LAT, Ajello et al. 2014 ) onboard the Fermi satellite, dozens of such events have been observed, some of them lasting for almost a day (Share et al. 2018; Omodei et al. 2018; Allafort 2018; Winter et al. 2018; Ajello et al. 2021). These events are thought to be sustained by particles accelerated at the CME-driven shock, long after the end of the flare impulsive phase and hence named as sustained gamma-ray emission (SGRE, Plotnikov et al. 2017; Jin et al. 2018; Kahler et al. 2018; Gopalswamy et al. 2018a; 2019). Emphasizing the occurrence of gamma-ray emission in the post-impulsive-phase, the events are also called "late phase >100 MeV γ-ray emission (LPGRE)" by some (Share et al. 2018; Kouloumvakos et al. 2020). In this paper we use the term SGRE.

One of the primary implications of the shock mechanisms is that the gamma-ray source is spatially extended. This was first shown to be the case when gamma-ray line emission was observed from the disk due to an eruption occurring ~10⁰ behind the limb (Cliver 1993; Vestrand and Forrest 1993; 1994). Since gamma-ray emission occurs deep in the chromosphere, it cannot reach an Earth observer from behind the limb. Therefore, the accelerated protons need to precipitate on the frontside limb, which can be enabled by the CME-driven shock extending to the frontside. Fermi/LAT observed three SGRE events from behind-the-limb eruptions (Pesce-Rollins et al. 2015a,b; Ackermann et al. 2017; Plotnikov et al. 2017; Jin et al. 2018; Grechnev et al. 2018; Hudson 2018; Gopalswamy et al. 2020). The event on 2014 September 1 occurred ~40⁰ behind the east limb, suggesting that protons are precipitating and producing gamma-rays at longitudinal distances up to 40⁰ from the eruption region. If the gamma-ray source is symmetric in longitude with respect to the eruption location, we infer a longitudinal extent of ~80⁰. The large spatial extent of the SGRE source is consistent with the large extent of the shock surrounding the CME flux rope (Jin et al. 2018; Gopalswamy et al. 2020). Three-dimensional reconstruction of the CME flux rope and the shock driven by it during the 2014 September 1 event has shown that the gamma-ray emission originated in the shock sheath intersecting the solar surface at the limb (Gopalswamy et al. 2020). Shock-accelerated particles diffuse back to the solar surface along open magnetic field lines that thread the shock nose in the outskirts of the CME flux rope to produce SGRE (see Gopalswamy et al. 2020). The large source extent implies that the observed gamma-ray flux from sources closer to the limb is likely underestimated because part of the source is behind the limb. In contrast to the SGRE source, the post-eruption arcade (PEA) size is relatively small. E.g., the average and maximum sizes of PEAs from solar cycle 23 are 90,000 and 200,000 km, respectively (Yashiro et al. 2013). The average size corresponds to 7º on the Sun, an order of magnitude smaller than the SGRE source size inferred from the 2014 September 1 event.

The shock idea has been bolstered by the recent result that the SGRE duration ($D_g$) is correlated with the duration of the associated type II radio burst ($D_{II}$) for all SGRE events with duration >3 hr: $D_g = (1.0 \pm 0.2)D_{II} + (0.1 \pm 2.1)$ (Gopalswamy et al. 2020). This close relationship confirms that the same shock accelerates high-energy protons required for SGRE and tens of keV electrons responsible for the radio burst (Gopalswamy et al. 2018a; 2019). Furthermore, the relation implies that when the shock is efficient in accelerating ~10 keV electrons, it is equally efficient in accelerating >300 MeV protons. Since it is well established that large gradual SEP events are due to CME-driven shocks (e.g., Reames 2013), there is strong support for the original suggestion by Forrest et al. (1985) that the >300 MeV protons producing gamma-rays via neutral pions are from the same source as the SEPs, viz., the CME-driven shock.

If the shock paradigm is correct, one expects a correlation between the number of high-energy protons inferred from the SEP event ($N_{SEP}$) and that ($N_g$) derived from the observed gamma-ray flux. Such a test was recently performed by De Nolfo et al. (2019a,b) using a set of 14 SGRE events observed by Fermi/LAT that had overlapping SEP observations from the Payload for Matter-Antimatter Exploration and Light Nuclei Astrophysics (PAMELA, Adriani et a;. 2014). The authors used >500 MeV proton numbers and found that $N_{SEP}$ and $N_g$ are uncorrelated with the ratio $N_{SEP}/N_g$ scattered over more than five orders of magnitude from ~$7.8\times10^{-4}$ to ~$5.0\times10^{2}$. A much smaller range (0.0015 to 0.5) was reported by Share et al. (2018), who used only 8 events. Based on the lack of $N_{SEP} - N_g$ correlation, de Nolfo et al. (2019a,b) concluded that gamma-ray emission is probably not due to protons diffusing back to the Sun from CME-driven shocks. Given that the CME speed in SGRE events is similar to that of ground level enhancement (GLE) events and correlated with SGRE fluence (Gopalswamy et al. 2019), the lack of correlation between $N_{SEP}$ and $N_g$ is puzzling. This contradiction motivated us to revisit the $N_{SEP} - N_g$ correlation. Based on systematic patterns found in the $N_{SEP} - N_g$ correlation plot, we show that (i) the gamma-ray flux is underestimated in SGRE events occurring closer to the limb resulting in an underestimation of $N_g$, and (ii) $N_{SEP}$ is underestimated in SGRE events occurring at latitudes far exceeding 13º because the highest energy particles are accelerated close to the nose that is not connected to an Earth observer (Gopalswamy et al. 2013; Gopalswamy and Mäkelä, 2014). Correcting for these two effects removes the apparent lack of $N_{SEP} - N_g$ correlation reported in previous studies.

## 2. SGRE Event List and Analysis

We consider the 14 Fermi/LAT >100 MeV SGRE events that had simultaneous PAMELA SEP observations in the energy range >80 MeV (de Nolfo et al. 2019a,b). In particular, we are interested in the number of >500 MeV protons ($N_{SEP}$) inferred from the associated SEP event and that ($N_g$) derived from the >100 MeV gamma-ray flux observed by Fermi/LAT. De Nolfo et al. (2019a) estimated $N_{SEP}$ from PAMELA data and used $N_g$ previously obtained by Share et al. (2018) from the Fermi/LAT data.

Figure 1 shows the $N_{SEP} - N_g$ scatter plot for the 14 gamma-ray events. A similar plot was first made by Share et al. (2018) for a subset of 8 events. The dates and heliographic coordinates of the underlying eruptions are noted on the plot. As pointed out by de Nolfo et al. (2019a,b), the $N_{SEP} - N_g$ correlation is not significant because the correlation coefficient $r = 0.43$ is smaller than the Pearson's critical correlation coefficient $r_c = 0.46$ with $p = 0.05$. This conclusion was also reached by Share et al. (2018) who found $r = 0.3$ for the 8 events. On the basis of the lack of correlation, de Nolfo et al. (2019a) concluded that SGRE is probably not due to protons diffusing back to the Sun from CME-driven shocks. A closer look at the heliographic coordinates of the eruptions in Fig. 1 indicates a systematic pattern. (i) Events near the limb (central meridian distance, CMD >60º) are generally located below the regression line and at the right extreme of the cluster of data points. (ii) Events at higher latitudes are located above the regression line, occupying the left extreme of the cluster of data points; these latitudes are high compared to the typical latitude (~13º) of solar eruptions that result in GLE events (Gopalswamy et al. 2013; Gopalswamy and Mäkelä 2014). One of the higher-latitude events (2012 January 27) is also a limb event (W71). As noted in the introduction, $N_g$ is underestimated in SGRE events occurring close to the limb because of partial occultation of the extended gamma-ray source, so the data points are expected to move up in the plot when corrected. $N_{SEP}$ is underestimated in SGRE events occurring at large ecliptic latitudes because the highest energy particles are accelerated only near the shock nose, so the data points are expected to move to the right after corrections. While de Nolfo et al. (2019a,b) attempted to correct for the latitude effect, they used large widths (in the range 36.9º to 44.5º with an average of 40.7º for the five events that need latitude correction in Table 1) for the particle distribution, so the correction was not significant. We show that the energy dependence of the Gaussian width is important in making an accurate correction to $N_{SEP}$ for events occurring at large ecliptic latitudes.

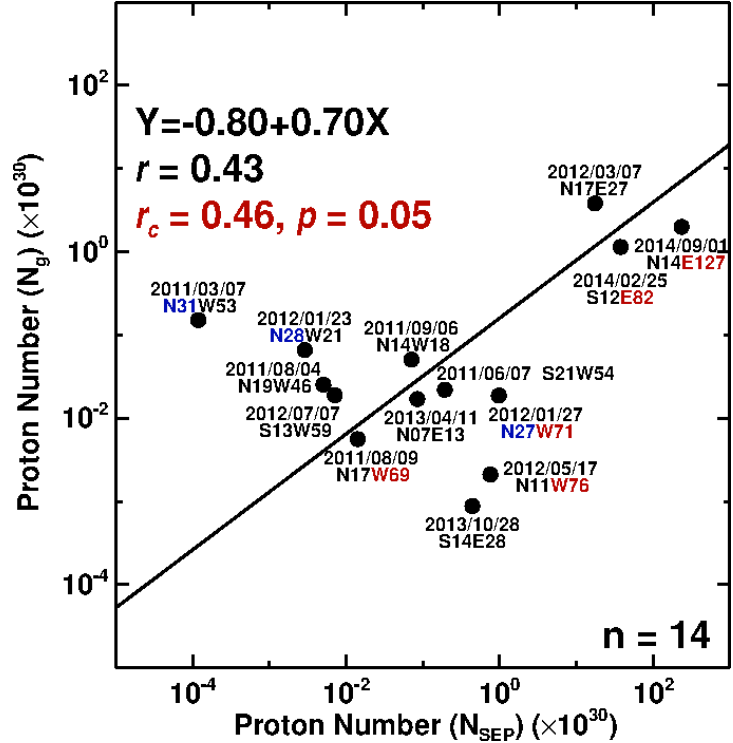

**Figure 1**. $N_{SEP} - N_g$ scatter plot obtained using the original data reported in de Nolfo et al. (2019a) for the 14 Fermi/LAT gamma-ray events. The event dates and the heliographic coordinates of the eruption regions are marked. The longitudes of events with central meridian distance ≥60º are marked in red. Events occurring at latitudes (poleward of ~25º) are marked in blue. The correlation coefficient ($r$), the Pearson's critical correlation coefficient ($r_c$), and the probability $p$ that the observed correlation is by chance are noted on the plot along with the regression equation.

In order to make these corrections, we have listed the relevant information on solar eruptions associated with the 14 events in Table 1. The information includes CME kinematic properties, eruption location, soft X-ray flare size, and the 10-100 MeV fluence spectral index of the associated SEP event.

The date and time of the SGRE events are listed in the first two columns. One of the distinguishing characteristics of CMEs associated with SGRE is that all underlying CMEs are halos as observed from Sun-Earth L1 Lagrange point by the Large Angle and Spectrometric Coronagraph (LASCO, Brueckner et al. 1995) on board the Solar and Heliospheric Observatory (SOHO). Halo CMEs are faster and wider on the average and hence represent an energetic CME population (Gopalswamy et al. 2010). For halo CMEs there is no central position angle, so they are characterized by the position angle at which the outer edge of the CME moves the fastest in the sky plane. This is the "measurement position angle" (MPA) listed in column 3. Most of the CMEs in Table 1 have been observed from multiple vantage points by coronagraphs on board SOHO and the Solar Terrestrial Relations Observatory (STEREO, Howard et al. 2008), so we were able to use the graduated cylindrical shell (GCS, Thernisien, 2011) model to get the three-dimensional (3-D) speed, acceleration, and direction of CMEs (see, e.g., Gopalswamy et al. 2014). The CME nose direction obtained from the GCS model is generally consistent with the MPA. Typically, the 3-D CME acceleration peaks close to the Sun, ~1.5 Rs, while the speed reaches a peak ($V_{pk}$) around 2.5 Rs (see e.g., Bein et al. 2011).  After attaining the peak, the CME speed slowly decreases owing to the aerodynamic drag. $V_{pk}$ and the initial CME acceleration ($a_i$) are listed in columns 5 and 6, respectively. If  $a_i$ is high (>1 km s$^{-2}$), the 10 – 100 MeV fluence spectrum of the associated SEP events ($F \sim E^{-\alpha}$) is hard (e.g., GLE events, the spectral index α ~ 2.61). On the other hand, if $a_i$ is small (a fraction of 1 km s$^{-2}$), which is typical of eruptions from filament regions outside active regions, the SEP spectrum is soft (α ~ 4.89, Gopalswamy et al. 2015; 2016). Other non-GLE SEP events have intermediate $a_i$ and spectral hardness. Gopalswamy et al. (2018b) reported that there is a group of events that have high $a_i$ (>5 km s$^{-2}$) but the SEP spectra are soft. In all these events, the CME nose is at an ecliptic distance much larger than the average distance (13º) in GLE events.  We have listed the peak $a_i$ obtained from GCS fit to the CMEs in column 5 and the 10-100 MeV fluence spectral index α in column 6. The peak $a_i$ are generally larger than the average initial acceleration computed from flare rise time and the average CME speed (Zhang and Dere 2006; Gopalswamy et al. 2016). The GCS fit also gives the edge-on and face-on half widths of the fitted flux rope. Column 7 gives the face-on half width (FHW), which gives the maximum extent of the CME flux rope in the coronagraph field of view.  The heliographic location and soft X-ray size of the associated flare are given in columns 8 and 9, respectively. $N_{SEP0}$ from PAMELA data obtained by de Nolfo et al. (2019a) is listed in column 10  while the latitudinal correction factor ($C_{lat}$) is in column 11 with the lower ($C_{lat1}$) and upper ($C_{lat2}$) values of the uncertainty range of $C_{lat}$ in columns 12 and 13. Column 14 gives $N_{g0}$ inferred from SGRE events (Share et al. 2018; de Nolfo et al. 2019a). The subscript "0" is used to denote that the numbers are original data from Table 6 of de Nolfo et al. (2019a). The correction factor ($C_{lon}$) in column 15 accounts for the gamma-ray flux reduction in limb events, along with the lower ($C_{lon1}$) and upper ($C_{lon2}$) values of the $C_{lon}$ uncertainty range. The last column gives the ratio $R = N_{SEP}/N_g$ after applying the corrections (explained in sections 2.1 and 2.2).

Some key properties of CMEs associated with SGRE can be readily obtained from Table 1. The average of the peak speeds is 2232 ±313 km s$^{-1}$, confirming previous result that the CMEs are extremely fast as in GLE events. The face-on half width of the CMEs averages to ~61º, indicating that the CMEs are fast and wide and hence very energetic. The peak $a_i$ averages to 4.03 ± 1.56 km s$^{-2}$, which is also very high as in GLE events. The fluence spectral index α ranges from 2.01 to 5.24. The two highest α (softest spectrum) are 5.24 and 4.70, which correspond to the two eruptions with the largest ecliptic distance: 2012 January 23 and 2011 March 7, respectively. These α values are similar to those in well-connected filament eruption (FE) events in which the CMEs accelerate slowly and attain relatively low peak speed. However, the CME speed and acceleration in these two events are GLE-like. Thus, a combination of high initial acceleration of CMEs and a soft-spectrum SEP events is a good indicator of poor magnetic connectivity between the SEP source and the observer (Gopalswamy et al. 2017; 2018b). Column 3 of Table 1 shows that there are five events in which the MPA is more than 25° away from the ecliptic (shown underlined). In all these cases, the longitudinal connectivity is good, so we need to correct only for the large latitudes. In the 2013 October 28 event, α ~4, which may be due to poor longitudinal connectivity to an Earth observer (GOES). However, this event and other eastern events are well-connected to STEREO, which observed harder power law.

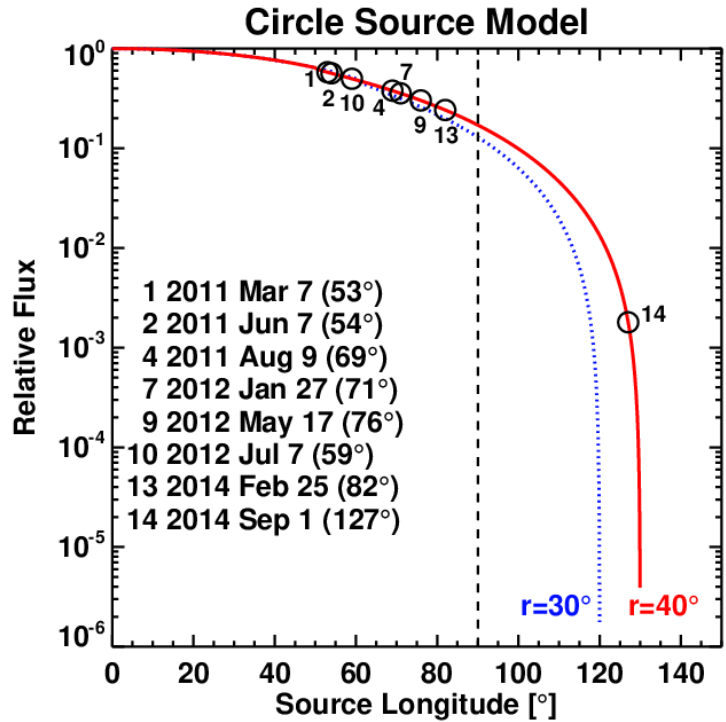

Figure 2. Observed gamma-ray flux relative to the emitted total flux as a function of source longitude. At the disk center, the gamma-ray source is assumed to be circular disk of radius 30° in blue and 40° in red emitting a flux of unity. The black open circles represent the gamma-ray flux from sources at longitude >50°. The observed gamma-ray flux is reduced for these sources by a factor ranging from ~0.60 for a central meridian distance of 53° to 0.0067 for a CMD of 127°. Therefore, the correction factor $C_{lon}$ ranges from 1.7 to 560. The vertical dashed line indicates the limb (CMD = 90°). In an eruption occurring at the solar limb, only ~20% of the emitted gamma-ray flux is observed by Fermi/LAT. The event numbers, dates, and source longitudes are listed on the plot (see also Table 1). The choice of a radius of 40° is based on the 2014 September 1 backside event. When the source starts crossing the limb, the observed gamma-ray flux is reduced, the reduction being larger for a source of smaller size.

## 2.1 Correcting for the longitude effect of SGRE sources

In order to correct for the longitude effect, we assume the gamma-ray source to be a circular disk with a radius of 40° (heliographic), equivalent to ~0.6 Rs. We use this number based on the fact the 2014 September 1 event originated ~40⁰ behind the limb, yet the gamma-ray emission was observed on the frontside at the limb. Individual events may have different longitudinal extents, but since the SGRE-associated CMEs constitute the fastest and widest of CME populations, we take this extent to be typical, consistent with the large face-on widths listed in Table 1.    As the CMD of eruptions increases, the gamma-ray flux decreases by cos φ, where φ is the CMD. When a segment of the circle goes behind the limb, we reduce the source area by subtracting the area of the behind-the-limb segment from the area of the full circle and use the mid-longitude of the on-disk segment as a nominal CMD.  We normalize the flux to be 1 at φ = 0⁰, so the inverse of the flux reduction gives the correction factor $C_{lon}$.

The normalized gamma-ray flux as a function of source longitude is plotted in Fig. 2. We see that the observed flux decreases as the source approaches the limb and starts to disappear behind the limb. We can correct for the resulting underestimation by multiplying $N_{g0}$ by the inverse of the relative flux ($C_{lon}$). The correction factor varies from 1.7 for a source at φ = 53⁰ to ~560 for the backside eruption at φ =127⁰. For source longitude <50⁰, there is no occultation of the limb ward section of the gamma-ray source, so the full area is visible. The curve corresponding to a source radius of 30⁰ shows that the flux reduction is deeper at larger φ, especially at the limb and beyond (see Fig. 2). Note that no gamma-ray flux would be observed from the 2014 September 1 backside event if the source radius were 30⁰. Since we do not know the actual extent of the gamma-ray source in each event, we varied the source radius from 20⁰ to 60⁰ and computed the uncertainty range of $C_{lon}$ bounded by $C_{lon1}$ to $C_{lon2}$ as listed in Table 1.

The longitudinal correction arising from the extended nature of the SGRE source is straightforward.  We assumed a circular disk whose radius is constrained by the 2014 September 1 backside eruption that produced a frontside SGRE signal ~40⁰ away from the eruption site.  It is possible that the actual source structure may be different from a circle. One may have to develop a sophisticated 3-D source model to further refine the results. In the simple geometric model, we consider only the flux reduction due to partial source occultation.  We do not consider the effect of the gamma-ray spectral shape and hence the dependence on the derived proton index. Share et al. (2018) computed the proton power law index from the fit to the gamma-ray data (their table 3).  The range of spectral indices is the same for events requiring/not requiring longitudinal correction.

## 2.2 Correcting for the latitude effect

In order to correct for the poor latitudinal connectivity, we need to consider the spectral steepness of the SEP event compared to the case when the source is within ~13⁰ from the ecliptic. For simplicity, we assume that the fluence spectral index is ~2 at the CME/shock nose, so the spectrum will be hard if the nose is well connected to the observer.  The basis for this assumption stems from the fact that well-connected GLE events have an average fluence spectral index of ~ 2. This has been shown to be the case from a study of GLE events from solar cycles 23 and 24 (Gopalswamy et al. 2016; 2018b). If an observer is not well connected to the shock nose, few high energy particles are detected, resulting in a soft spectrum. Even though the fluence spectral index computed by Gopalswamy et al. (2016) is for the energy range 10-100 MeV, we assume that the spectral index applies to GeV energies (i.e., no rollover at high energies). This is a good approximation only for some events such as the one on 2012 March 7. The effect of relaxing this assumption on the latitude correction will be discussed in section 4. We normalize the 1 GeV intensity to unity, which makes the intensity at 100 MeV and 500 MeV to be $10^2$ and 4, respectively for $\alpha = 2$.

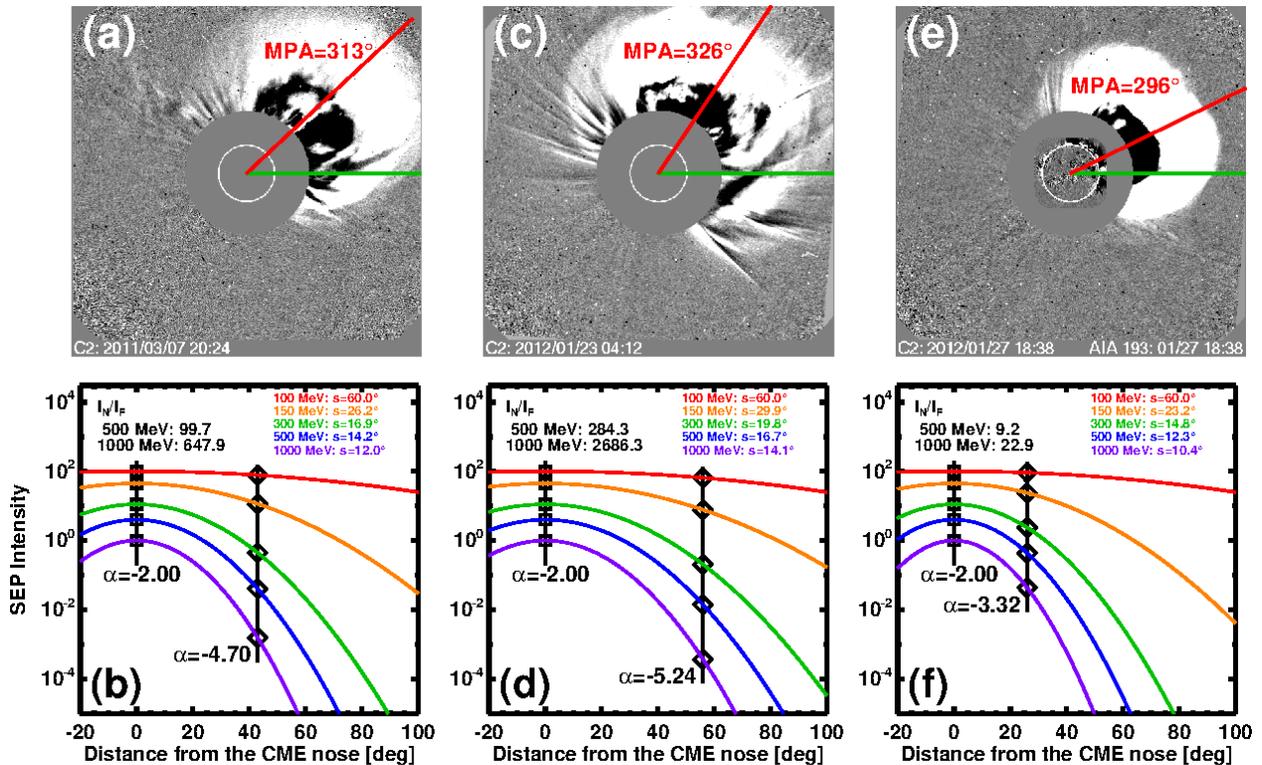

Figure 3. Three CMEs observed by SOHO/LASCO with their noses at relatively large ecliptic distances: 2011 March 7 (a), 2012 January 23 (c), and 2012 January 27 (e), with the corresponding behavior of the SEP intensity as a function of the angular distance $w$ from the CME/shock nose in (b), (d), and (f). In the top row, the ecliptic (red line) and the measurement position angle (MPA) corresponding to the CME nose (green line) are marked. The Gaussian widths are energy dependent and are constrained by the fluence spectral index $\alpha$ and the width of the 100 MeV Gaussian set as 60º. When an observer is connected to the nose, a hard spectrum is detected (spectral index $\alpha \sim 2$). When the observer is not connected to the nose, a steeper spectrum results depending on the distance ($\lambda$) from the CME nose to the observer in the ecliptic. The ratio of the nose intensity ($I_N$) to the flank intensity ($I_F$) is given on the plots for 1 GeV and 500 MeV. Also shown are the Gaussian widths ($s$) at various energies determined from Eq. (1).

To estimate the intensity at the nose based on the flank observations, we need to know how the intensity falls off from the nose. If we assume a Gaussian falloff, one can determine the intensity at the flank for various energies and Gaussian widths. The intensity ($I_E$) at energy $E$ is given by a Gaussian in $w$, the angle from the CME nose:

$$I_E = A_E e^{-\frac{1}{2}(w/s_E)^2}. \quad (1)$$

Here the unknowns are $A_E$, a constant and $s_E$ the Gaussian width at each energy. For a given energy, Eq. (1) can be written as two simultaneous equations, one for the nose ($w = 0$), and the other for the flank ($w = \lambda$, the observer position for each event) from the spectra. Once we know $s_E$, we can determine the nose intensity ($I_N$) from the observed flank intensity ($I_F$), and hence the correction factor $C_{lat} = I_N/I_F$ at each energy. The steps are as follows. (1) Normalize the 1-GeV intensity ($I_E = 1$) and assume $\alpha = 2$ at the nose, so we know $I_E$ at lower energies. We choose 100 MeV as the lowest energy to be similar to the PAMELA energy range (>80 MeV) and well below the energy range of protons that result in SGRE. (2) Impose the condition that the Gaussian width at the lowest energy (100 MeV) is 60º, so we know $I_E$ at the nose and flank at 100 MeV from Eq. (1). This condition is consistent with the observation that all the SGRE CMEs are full halos indicating that the underlying shocks are well extended (more later on the 60º width constraint). (3) Obtain the flank intensities at higher energies from the 100 MeV flank intensity using the observed spectrum (e.g., $\alpha = 4.70$ for the 2011 March 7 event with $\lambda = 43$º). (4) Now that the nose and flank intensities are known at all energies in the range 100 MeV – 1 GeV, determine the Gaussian width $s_E$ at each energy. Figure 3 shows the resulting energy-dependent Gaussians for three high-latitude events at several particle energies: 1 GeV, 500 MeV, 300 MeV, 150 MeV, and 100 MeV.

For the 2011 March 7 event, we get $s_E \sim 12$º at 1 GeV and $\sim 14$º at 500 MeV. The correction factor $C_{lat}$ needed to bring $I_F$ to the nose level $I_N$ is the ratio $I_N/I_F$. Figure 3b shows that $C_{lat} \sim 648$ at 1 GeV and $\sim 100$ at 500 MeV. In the soft-spectrum event of 2012 January 23 ($\alpha = 5.24$; $\lambda = 56$º, the 1-GeV and 500-MeV widths are $\sim 14.1$º and $\sim 16.7$º, respectively resulting in $C_{lat} = \sim 2686$ at 1 GeV and $\sim 284$ at 500 MeV. Even though the 2012 January 27 CME originated from the same active region as the January 23 CME the fastest moving section of the leading edge is at a position angle of 296º ($\lambda = 26$º, see Fig. 3e). The spectrum is relatively hard ($\alpha = 3.62$), therefore, $C_{lat}$ is smaller: $\sim 23$ (1 GeV) and $\sim 9$ (500 MeV). The Gaussian widths are also smaller: 10.4º (1 GeV) and 12.3º (500 MeV). The range of 1-GeV Gaussian widths (10º – 17º) is consistent with the previous empirical findings that GLE events have a typical ecliptic distance of $\sim 13$º.

We repeated the $C_{lat}$ calculations by changing the 100-MeV Gaussian width to 90º and 120º. The resulting correction factors are very similar to the 60º case differing by a small factor varying between 1.06 and 1.39. Therefore, the 100 MeV width of 60º is quite reasonable and allows for even wider shock surface at lower energies. In order to estimate the effect assuming the Gaussian width at 100 MeV, we varied the width from 30º to 60º and computed the range of $C_{lat}$. The minimum ($C_{lat1}$) and maximum ($C_{lat2}$) values are shown in Table 1.

As can be seen in figures 3b,d, and f, the Gaussian widths ($s$) decrease with increasing energy. The width dependence on energy can be represented by a polynomial:

$$s(E) = B_n \sum_n \left(\frac{E}{100}\right)^{-n}, (2)$$

where $B_n$ are the coefficients with n = 0, 1, 2, …. The fit is reasonable even for $n = 1$: for the 2011 March 7 event, $B_0 = 7.14$ and $B_1 = 33.38$, giving the width at 500 MeV as 13.8° similar to the width in Fig. 3b (14.2°). A first order polynomial in $E$/100 is a reasonable representation of the energy-dependent widths in the other two events as well. For the five events that need latitudinal correction, de Nolfo et al. (2019a) used Gaussian widths in the range 36.9° to 44.5° (see their table 6), which are at least 3 times larger than the widths we obtained. Such large widths are acceptable only for energies below 150 MeV in order to satisfy the soft spectrum (see Fig. 3); using such a large width at higher energies results in a negligible correction ($C_{lat}$ ~1). Therefore, we apply the latitude correction to $N_{SEP0}$ values given in Table 6 of de Nolfo et al. (2019a), noting that their correction mainly takes care of longitudinal connectivity.

## 3. The $N_{SEP} - N_g$ correlation

We applied the latitude and longitude corrections ($C_{lat}$, $C_{lon}$) to the original numbers $N_{SEP0}$ and $N_{g0}$ in Table 1. The resulting $N_{SEP} - N_g$ scatter plot is shown in Fig. 4a. We see that the rightmost and leftmost data points systematically move up and to the right, respectively. The length of the arrows indicates the extent of the corrections. The four data points that do not need corrections (red filled circles in Fig. 4a) are bracketed by the leftmost ones requiring $C_{lat}$ and the rightmost ones requiring $C_{lon}$. Three data points need both latitude and longitude corrections (slanted arrows in Fig. 4a). We see that the corrections greatly improve the $N_{SEP} - N_g$ correlation with $r = 0.71$, which is 65% higher than the uncorrected case ($r = 0.43$, see Fig. 1). The correlation is highly significant because $r$ greatly exceeds the Pearson's correlation coefficient ($r_c = 0.46$ for $p = 0.05$; $r_c = 0.66$ for $p = 0.005$), $p$ being the probability that the obtained correlation is by chance. The slope of the regression line is close to unity, further confirming the close relation between $N_{SEP}$ and $N_g$.

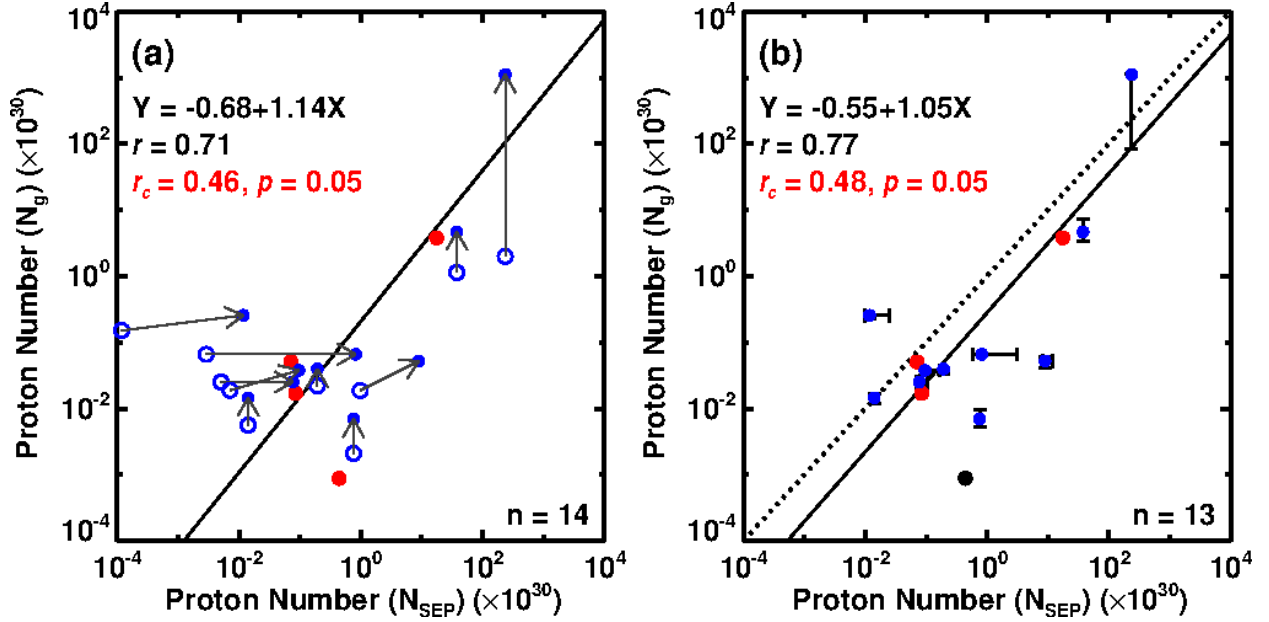

Figure 4. (a) $N_{SEP} - N_g$ scatter plot showing the corrections applied. The arrows point to the corrected data points (filled blue circles) from the corresponding original data points (open blue circles). The filled red circles denote data points that do not need a correction. The regression line ($N_g = -0.72 + 1.09N_{SEP}$) is the best-fit lines to the corrected data points. Note that the high-latitude points shift systematically to the right, while the high-CMD points shift systematically upward from the pre-correction positions. The correlation coefficient ($r = 0.71$) is much larger than the Pearson's critical correlation ($r_c = 0.46$) indicating statistically significant correlation. The probability $p$ that the observed correlation by chance is 0.05. The correlation remains significant ($r_c = 0.66$) even for $p = 0.005$. (b) The same scatter plot as in (a) but excludes the 2013 October 28 event (black filled circle) in the correlation. The correlation coefficient ($r = 0.77$) increases slightly and is much larger than the Pearson's critical correlation ($r_c = 0.48$, $p = 0.05$; $r_c = 0.68$, $p = 0.005$). The regression line becomes ($N_g = -0.59 + 0.99N_{SEP}$). The dotted line represents the bisector ($N_g = N_{SEP}$). By varying the gamma-ray source radius from 20º to 60º in steps of 5º, we obtained the uncertainty range in $C_{lon}$, which is then converted into the uncertainty range in $N_g$. By varying the 100 MeV Gaussian width from 30º to 120º we obtained the uncertainty range in $C_{lat}$, which is then converted into the uncertainty range in $N_{SEP}$.

The 2013 October 28 is an outlier with the lowest $N_g$. Table 1 shows that this event had the lowest speed (1251 km s$^{-1}$) and acceleration (0.82 km s$^{-2}$) among the 14 events. Based on the time profile of the gamma-ray emission in this event, Share et al. (2018) concluded that they are not able to determine whether the observed emission is just the tail of impulsive-phase radiation or a distinct late-phase component. If we exclude this event, the correlation improves slightly ($r = 0.77$, $r_c = 0.48$, p = 0.05 and $r_c = 0.68$, p = 0.005). Thus, the $N_{SEP} - N_g$ correlation is high with and without the 2013 October 28 event. The $N_g = N_{SEP}$ line is above the regression line ($N_g = -0.59 + 0.99N_{SEP}$) by a factor of ~3.9. i.e., $N_{SEP}$ is ~3.9 times $N_g$. We also note that only two data points are above the $N_g = N_{SEP}$ line, indicating that $N_{SEP} > N_g$ in most cases. The $N_{SEP}/N_g$ ratio varies from 0.05 to 178, with a median of 4.6 for the 13 events in Fig. 4b. If the 2013 October 28 event is included, the upper value increases to ~502, giving a median $N_{SEP}/N_g$ ratio of 4.8. In the uncorrected case, the $N_{SEP}/N_g$ ratio varied from 0.00078 to ~500 (de Nolfo et al. 2019a), indicating almost two orders of magnitude improvement after the correction. The propagation conditions are expected to be very different for particles propagating away and toward the Sun, so some amount of variability is expected (see Dalla et al. 2020). Mirroring near the Sun may create a situation $N_{SEP} > N_g$, but when eruptions are closely spaced, there may be mirroring toward the Sun (Gopalswamy et al. 2019) setting up the possibility of reversing this inequality.

## 4. Discussion and Conclusions

The primary result of this paper is that the number of >500 MeV protons ($N_g$) deduced from the Fermi/LAT >100 MeV SGRE is significantly correlated with the number of SEP protons at 1 au ($N_{SEP}$) estimated from PAMELA SEP measurements. The lack of correlation previously reported has been shown to be due to (i) a systematic underestimate of $N_g$ in events originating close to the limb due to the spatially-extended nature of SGRE under the shock paradigm, and (ii) a systematic underestimate of $N_{SEP}$ in events originating at higher latitudes as a consequence of the energy-dependent width of SEP source on the shock surface. The systematic effect is clear from the scatter plot in Fig. 1, in which the three groups of events occupy distinct regions of the plot: high-latitude events to the left, high-CMD events to the right, and the disk-center events in the middle. CME-driven shocks as the source of >300 MeV protons inferred from the close correlation between $N_{SEP}$ and $N_g$ is consistent with other known correlations: (i) the SGRE duration is linearly related to the duration and ending frequency of the associated interplanetary type II radio burst, indicating the underlying proton and electron accelerations are by the same shock over the duration that the shock remains potent (Gopalswamy et al. 2018a; 2019), (ii) the SGRE fluence is significantly correlated with CME speed (Gopalswamy et al. 2019), and (iii) the evolution of shock parameters on field lines returning to the visible disk are consistent with the time history of >100 MeV gamma-rays in the 2017 September 10 SGRE event (Kouloumvakos et al. 2020).

Since CMEs involved in SGRE have properties similar to the ones causing GLE events, we assumed the particle spectrum to be hard ($\alpha = 2.0$) at the CME/shock nose with a normalized intensity of unity at 1 GeV. Given $\alpha$ and the 1-GeV intensity at the nose, the nose intensity at various energies down to 100 MeV are known from the power law. Constrained by the observed SEP spectral index at a connection angle determined from coronagraph observations, we determined the flank SEP intensity and the Gaussian width at various energies. We find that the Gaussian width at 1 GeV is in the range $10^o – 14^o$ similar to the observed average ecliptic distance (~13$^o$) in well-connected GLE events. The Gaussian width increases toward lower energies attaining ~40$^o$ at energies <150 MeV. Interestingly, the two well-connected, hard-spectrum events on 2011 August 9 (N17W69, $\alpha = 2.50$) and 2017 May 17 (N11W76, $\alpha = 2.48$) that do not need latitudinal correction, have been reported to have Gaussian widths of 21$^o$ and 20.8⁰, respectively in de Nolfo et al. (2019a). The widths computed by de Nolfo et al. (2019a) are consistent with our result that well-connected events have nose connectivity to the observer. Finally, we also checked the effect of assuming $\alpha = 2.0$ at the nose. If the nose spectral index is changed to the observed average value ($\alpha \sim 2.5$) instead of the ideal case of $\alpha = 2.0$, $C_{lat}$ increased by a factor of $\sim 2$ with no change in the correlation coefficients.

Now we consider the effect of high energy rollover in SEP spectra. Bruno et al. (2018) have shown that many of the SEP events detected by PAMELA show a high energy rollover following the Ellison-Ramaty (E-R) spectrum: $F(\alpha) \sim E^{-\alpha} \exp(-E/E_0)$, where $E_0$ is the rollover energy (Ellison and Ramaty, 1985). The fluence spectra of ten of the 14 events in our Table 1 have been reported by Bruno et al. (2018) with E-R fits. The ten include three of the four high-latitude events: 2012 January 23, 2012 January 27, and 2012 July 7. The power-law part of the E-R spectra is similar to that from GOES data reported in Gopalswamy et al. (2016): $\alpha = 5.33, 3.36,$ and 2.55 (Bruno et al. 2018) vs. 5.24, 3.32, and 3.5 (Gopalswamy et al. 2016). When we use the E-R spectrum instead of the power law in Fig. 3, we get $C_{lat} = 328.6, 9.8,$ and 2.9 for the three events. These values are not too different from those in the power law case: 284.3, 9.2, and 13.5, respectively. When we use the new $C_{lat}$ values, the $N_{SEP} - N_g$ correlation coefficients remain the same as in Fig. 4 (0.71 for all 14 events and 0.77 when the 2013 October 28 event is excluded).

The main conclusions of this work can be summarized as follows:

1. The number of high-energy protons at 1 au derived from the SEP events is highly correlated with that inferred from the observed gamma-ray flux in SGRE events. The high correlation ($r = 0.77$) is consistent with the shock paradigm.

2. The number of >500 MeV protons derived from the SEP events is higher than that derived from the SGRE event by a factor of $\sim 4$.

3. The correlation is unchanged when an Ellison-Ramaty spectrum is used instead of a power law in the correction factor for the number of >500 MeV protons inferred from the SEP event.

4. The distribution of accelerated particles can be represented as a series of energy-dependent Gaussians centered at the shock nose with the width decreasing with increasing energy following a low-order polynomial in energy.

5. The highest energy (GeV) particles are accelerated near the nose region with a Gaussian width in the range 10⁰ – 17⁰. The energy-dependent width thus explains why the average ecliptic distance is $\sim 13$⁰ in eruptions that result in a GLE.

6. The peak speed ($2242 \pm 313$ km s⁻¹) and initial acceleration ($4.03 \pm 1.56$ km s⁻²) of the underlying CMES for the events considered in this work are similar to those in GLE events. This means particles are accelerated to GeV energies in SGRE events, guaranteeing the presence of >300 MeV particles above the shock nose.

This work benefitted greatly from the open data policy of NASA in using SDO, SOHO, STEREO, and Wind data. STEREO is a mission in NASA's Solar Terrestrial Probes program. SOHO is a project of international collaboration between ESA and NASA. The work of N.G., S.Y., and S.A. was supported by NASA/LWS program. H.X. was partially supported by NASA grant NNX15AB70G.

**References**

Ackermann, M., Allafort, A., Baldini, L., et al. 2017, ApJ, 835, 219

Adriani, O., Barbarino, G. C., Bazilevskaya, G. A., et al. 2014, PhR, 544, 323

Ajello M, Albert A, Allafort A, et al. 2014 *ApJ* **789** 20

Ajello, M., Baldini, L., Bastieri, D. 2021 ApJ, 252, 13

Allafort, A. J., High-energy gamma-ray observations of solar flares with the Fermi large area telescope, PhD thesis, Stanford University, 2018.

Bein, B. M., Berkebile-Stoiser, S., Veronig, A. M., et al. 2011, ApJ, 738, 191

Brueckner, G.E., Howard, R.A., Koomen, M.J., et al. 1995, SoPh, 162, 357
Bruno, A., Bazilevskaya, G. A., Boezio, M., et al. 2018, ApJ, 862, 97

Chupp, E. L., & Ryan, J. M. 2009, RAA, 9, 11

Cliver, E. W., Kahler, S. W., & Vestrand, W. T. 1993, ICRC (Calgary), 3, 91

Dalla, S., de Nolfo, G. A., Bruno, A., et al. 2020, A&A, 639, 105

De Nolfo, G. A., Bruno, A., Ryan, J. M., et al. 2019a, ApJ, 879, 90

De Nolfo, G. A., Bruno, A., Ryan, J. M., et al. 2019b, *36th International Cosmic Ray Conf.* 358, 1073
Ellison, D. C., & Ramaty, R. 1985, ApJ, 298, 400

Forrest, D. J., Vestrand, W. T., Chupp, E. L., et al. 1985, ICRC, 4, 146

Gopalswamy, N., Yashiro, S., Michalek, G., et al. 2010, SunGeo, 5, 7

Gopalswamy, N., Xie, H., Akiyama, S., et al. 2013, ApJL, 765, L30

Gopalswamy, N., & Mäkelä, P. 2014, in ASP Conf. Ser. 484, Outstanding Problems in Heliophysics: from Coronal Heating to the Edge of the Heliosphere, ed. Q. Hu & G. Zank (San Francisco, CA: ASP), 63
Gopalswamy, N., Xie, H., Akiyama, S. et al. 2014, EPS, 66, 104

Gopalswamy, N., Mäkelä, P., Akiyama, S., et al. 2015, ApJ, 806, 8

Gopalswamy, N., Yashiro, S., Thakur, N. et al. 2016, ApJ, 833, 216.

Gopalswamy, N., Mäkelä, P., Yashiro, S., et al. 2017, JPhCS, 900, 012009

Gopalswamy, N., Mäkelä, P., Yashiro, S., et al. 2018a, ApJL, 868, L19

Gopalswamy, N., Yashiro, S., Mäkelä, P., et al. 2018b, ApJ, 863, L39

Gopalswamy, N., Mäkelä, P., Yashiro, S. et al. 2019, JPhCS, 1332(1), 012004

Grechnev, V. V., Kiselev, V. I., Kashapova, L. K., et al. 2018, SoPh, 293, 133

Gopalswamy, N., Mäkelä, P., Yashiro, S., et al. 2020, SoPh, 295, 18

Howard, R. A., Moses, J. D., Vourlidas, A. et al. 2008, SSRv, 136, 67

Hudson, H. S. 2018, in IAU Symp. 335, Space Weather of the Heliosphere: Processes and
        Forecasts, ed. C. Foullon & O. E. Malandraki (Cambridge: Cambridge Univ. Press), 49

Jin, M., Petrosian, V., Liu, W., et al. 2018, ApJ, 867, 122

Kahler, S. W., Cliver, E. W., Kazachenko, M. 2018, ApJ, 868, 81

Kouloumvakos, A., Rouillard, A. P., Share, G. H., et al. 2020, ApJ, 893, 76

Murphy, R.J., Dermer, C.D., Ramaty, R.: 1987, ApJ, 63, 721.

Omodei, N., Pesce-Rollins, M., Longo, F. et al. 2018, ApJ, 865, L7

Pesce-Rollins, M., Omodei, N., Petrosian, V. et al. 2015a, ApJL, 805, L15

Pesce-Rollins, M., Omodei, N., Petrosian, V., et al.. 2015b, 34th International Cosmic Ray
        Conf., 128.

Plotnikov, I., Rouillard, A. P., & Share, G. H. 2017, A&A, 608, A43

Reames, D. V. 2013, SSRv, 175, 53

Richardson, J. D., Wang, C., Kasper, J. C., Liu, Y. 2005, GRL, 32, L03S03

Ryan, J. M., & Lee, M. A. 1991, ApJ, 368, 316

Share, G. H., Murphy, R. J., Tolbert, A. K., et al. 2018, ApJ, 869, 182

Thernisien, A. 2011, ApJS, 194, 33

Vestrand W. T., & Forrest D. J. 1993, ApJL, 409, L69

Vestrand W. T., & Forrest D. J. 1994, AIP Conference Proceedings 294, 143

Winter, L. M., Bernstein, V., Omodei, N., et al. 2018, ApJ, 864, 39

Yashiro S., Gopalswamy N., Mäkelä P., Akiyama S. 2013, SoPh, 284, 5

Zhang, J. & Dere, K. P. 2006, ApJ, 649, 1100

Table 1. CME, Flare, and SEP properties of SGRE events

| EDATE | ETIME | MPA[b] | $V_{pk}$ | $a_i$ | $\alpha$ | FHW[k] | Location | Flare | $N_{SEP0}$ | $C_{lat}$ | $C_{lat1}$ | $C_{lat2}$ | $N_g$ | $C_{lon}$ | $C_{lon1}$ | $C_{lon2}$ | $R^l$ |
|---|---|---|---|---|---|---|---|---|---|---|---|---|---|---|---|---|---|
| 2011/03/07 | 20:00:05 | 313 | 2660 | 1.52 | 4.70c | 48 | N31W53 | M3.7 | 1.18e+26 | 99.7 | 82.2 | 215.4 | 1.51e+29 | 1.7 | 1.5 | 1.7 | 0.05 |
| 2011/06/07 | 06:49:12 | 250 | 1680 | 2.43 | 2.22c | 53 | S21W54 | M2.5 | 1.93e+29 | 1 | 1 | 1 | 2.20e+28 | 1.8 | 1.5 | 1.8 | 4.87 |
| 2011/08/04 | 04:12:05 | 298 | 2475 | 5.81 | 3.62c | 58 | N19W36 | M9.3 | 5.04e+27 | 15.2 | 13.9 | 21.0 | 2.52e+28 | 1 | 1 | 1 | 3.04 |
| 2011/08/09 | 08:12:06 | 279 | 2496 | 8.16 | 2.50c | 57 | N17W69 | X6.9 | 1.41e+28 | 1 | 1 | 1 | 5.60e+27 | 2.6 | 2.1 | 3.0 | 0.97 |
| 2011/09/06 | 23:05:57 | 70 | 1474 | 1.59 | 3.22d | 65 | N14W18 | X2.1 | 7.09e+28 | 1 | 1 | 1 | 5.06e+28 | 1 | 1 | 1 | 1.40 |
| 2012/01/23 | 04:00:05 | 326 | 2158 | 6.27 | 5.24e | 88 | N28W21 | M8.8 | 2.89e+27 | 284.3 | 205.1 | 1050.3 | 6.60e+28 | 1 | 1 | 1 | 12.45 |
| 2012/01/27 | 18:27:52 | 296 | 2755 | 3.89 | 3.32c | 78 | N27W71 | X1.8 | 9.74e+29 | 9.2 | 8.6 | 12.2 | 1.87e+28 | 2.8 | 2.2 | 3.3 | 171.14 |
| 2012/03/07 | 00:24:06 | 57 | 2987 | 4.67 | 3.71e | 54 | N17E27 | X5.4 | 1.75e+31 | 1 | 1 | 1 | 3.81e+30 | 1 | 1 | 1 | 4.59 |
| 2012/05/17 | 01:45:05 | 261 | 1997 | 2.73 | 2.48e | 69 | N11W76 | M5.1 | 7.56e+29 | 1 | 1 | 1 | 2.10e+27 | 3.3 | 2.5 | 4.5 | 109.09 |
| 2012/07/07 | 23:24:06a | 233 | 2464 | 7.82 | 3.50f | 46 | S13W59 | X1.1 | 7.10e+27 | 13.5 | 11.7 | 23.9 | 1.89e+28 | 2.0 | 1.7 | 2.1 | 2.54 |
| 2013/04/11 | 07:24:06 | 85 | 1626 | 1.39 | 3.49g | 76 | N09E12 | M6.5 | 8.47e+28 | 1 | 1 | 1 | 1.70e+28 | 1 | 1 | 1 | 4.98 |
| 2013/10/28 | 15:36:05 | 86 | 1251 | 0.82 | 4.03h | 58 | S06E28 | M4.4 | 4.41e+29 | 1 | 1 | 1 | 8.80e+26 | 1 | 1 | 1 | 501.14 |
| 2014/02/25 | 01:25:50 | 73 | 2777 | 7.83 | 3.10i | 70 | S12E82 | X5.0 | 3.77e+31 | 1 | 1 | 1 | 1.14e+30 | 4.1 | 3.0 | 6.4 | 8.07 |
| 2014/09/01 | 11:12:05 | 65 | 2450 | 1.42 | 2.01j | 38 | N14E127 | X2.4 | 2.35e+32 | 1 | 1 | 1 | 1.99e+30 | 560 | 42 | 560 | 0.21 |

[a]Time corresponds to the previous day

[b]Measurement position angle (MPA), because all CMEs are halos (see https://cdaw.gsfc.nasa.gov/CME_list/halo/halo.html for details on the CMEs in Table 1).

[c]From Gopalswamy et al. (2016)

[d]From GOES data excluding 10 MeV data point (affected by the preceding event),

[e]Low-energy GOES channels are affected by energetic storm particle events, so index corresponds to the range 40-600 MeV. This is a compound event with two eruptions originating from the same active region. The second CME is also very fast with PA = 82º, so no latitude correction is needed.

[f]The spectral index was listed as 2.50 in Gopalswamy et al. (2016) due to a typographical error.

[g]10 MeV data point excluded because of a rollover at lower energies

[h]Intense emission from a previous event made it difficult to determine the onset and end times at low energies, so only 40 – 110 MeV data are used to get the spectral index. The fluence computation is also affected by the arrival of a shock and the occurrence of a halo CME the next day.

[i]ESP contribution at low energies, so only data points in the range 40-100 MeV were used.

[j]From Gopalswamy et al. (2020)

[k]Face-on half width (FCW) in degrees obtained from the GCS model fit to the white light CMEs

[l]The corrected ratio $N_{SEP}/N_g$ with $N_{SEP} = N_{SEP0} \times C_{lat}$ and $N_g = N_{g0} \times C_{lon}$. $N_{SEP0}$ and $N_{g0}$ are the original >500 MeV proton numbers listed in Table 6 of de Nolfo et al. (2019a)